\documentclass[12pt,a4paper]{article}
\usepackage{amssymb}
\newcommand{\np}{Nucl. Phys.\ }

\newcommand{\pr}{Phys. Rev.\ }

\newcommand{\lqcd}{\Lambda_{\rm QCD}}

\def\del{\partial}

\def\hat{\widehat}
\def\tilde{\widetilde}
\def\OSW{{\cal O}_{SW}}
\def\qbar{{\overline{q}}}
\def\slash#1{\mbox{$\not \!\! #1$}}

\def\rDslash{{\overrightarrow{\slash D}}}
\def\lDslash{{\overleftarrow{\slash D}}}

\def\spose#1{\hbox to 0pt{#1\hss}}
\def\ltapprox{\mathrel{\spose{\lower 3pt\hbox{$\mathchar"218$}}
 \raise 2.0pt\hbox{$\mathchar"13C$}}}
\def\gtapprox{\mathrel{\spose{\lower 3pt\hbox{$\mathchar"218$}}
 \raise 2.0pt\hbox{$\mathchar"13E$}}}
\def\inapprox{\mathrel{\spose{\lower 3pt\hbox{$\mathchar"218$}}
 \raise 2.0pt\hbox{$\mathchar"232$}}}

\begin{document}

\pagestyle{empty} 
\begin{flushright}
EDINBURGH 96/28 \\
ROME1-1159/97 \\
ROM2F-20/97 \\
SHEP 97-05 \\
UW/PT 96-33 \\
May 1997
\end{flushright}
\centerline{\LARGE{\bf Non-perturbative improvement of composite}}
\vskip 0.2cm
\centerline{\LARGE{\bf operators with Wilson fermions}}
\vskip 0.3cm
\vskip 1cm
\centerline{\bf{G.~Martinelli$^a$, G.C.~Rossi$^b$, C.T.~Sachrajda$^c$,
S.~Sharpe$^{a,d}$,}}
\vskip 0.2cm
\centerline{\bf{M.~Talevi$^e$ and M.~Testa$^{a}$}}
\vskip 0.3cm
\centerline{$^a$ Dip. di Fisica, Univ. di Roma ``La Sapienza'' and
INFN, Sezione di Roma,}
\centerline{P.le A. Moro 2, I-00185 Roma, Italy}
\smallskip
\centerline{$^b$ Dip. di Fisica, Univ. di Roma ``Tor Vergata''
and INFN, Sezione di Roma II,}
\centerline{Via della Ricerca Scientifica 1, I-00133 Roma, Italy}
\smallskip
\centerline{$^c$ Department of Physics, University of Southampton,
Southampton SO17 1BJ, UK}
\smallskip
\centerline{$^d$ Physics Dept., University of Washington,
Seattle WA 98195, USA}
\smallskip
\centerline{$^e$ Department of Physics \& Astronomy, University of Edinburgh}
\centerline{The King's Buildings, Edinburgh EH9 3JZ, UK}

\vskip 2cm
\centerline{\bf ABSTRACT}
\begin{quote}
{We propose a method to improve lattice operators composed of Wilson
fermions which allows the removal of all corrections of $O(a)$,
including those proportional to the quark mass, leaving only errors of
$O(a^2)$. The method exploits the fact that chiral symmetry is
restored at short distances. By imposing this requirement on
correlation functions of improved lattice operators at short
distances, the coefficients which appear in these operators can be
determined. The method is an extension of the improvement program of
the {\sc Alpha} collaboration, which, up to now, has only been
applicable in the chiral limit. The extension to quarks with non-zero
masses is particularly important for applications in heavy quark
physics.}
\end{quote}
\vfill

\newpage


\pagestyle{plain}
\setcounter{page}{1}

\section{Introduction}

An important source of errors in present lattice computations is the
use of a finite lattice spacing, $a$. A systematic method for reducing
discretization errors order by order in $a$ was proposed by
Symanzik~\cite{Symanzik} and developed in ref.~\cite{Luscher}. It
consists of modifying the action and operators by ``irrelevant'' terms,
chosen in such a way that the convergence to the continuum limit is
accelerated. For Wilson fermions, in the first implementations of this
procedure, the improvement coefficients were computed in perturbation
theory, leaving errors of $O( a g_0^{2})$ in physical
quantities~\cite{sw,heatlie}, where $g_0$ is the bare coupling constant. 
It turns out, however, that, for lattice
spacings used in present simulations, one-loop perturbation theory is
not sufficiently accurate for some quantities~\cite{bknonpert,jlqcd},
even when ``tadpole resummed"~\cite{lepagemack}. Recently, a method for
determining the improvement coefficients beyond perturbation theory has
been proposed and implemented, thus achieving full $O(a)$
improvement~\cite{Luscher2,Luscher3}~\footnote{ By full $O(a)$
improvement we mean that the remaining discretization errors are of
$O(a^2)$.}.  This method allows one to obtain the improved action and
the improved vector currents, but can only be applied to other
composite operators, including the axial currents, in the chiral limit.
We explain the reasons for this limitation in section~\ref{sec:wis}.
It is particularly significant for applications involving
heavy quarks, where $O(a m)$ errors may be large~\footnote{ By $m$ we
mean some choice of physical quark mass.  We do not need to pick a
specific definition in this paper.}. Important examples in the
phenomenology of $B$-mesons include the calculations of the leptonic
decay constant $f_B$, the form factors of semileptonic $B$ decays, the
$B$-parameters and the amplitudes of the radiative decay $B\to
K^*\gamma$. \par

In this paper we suggest a new method, based on the short-distance
behaviour of gauge-invariant correlation functions, which overcomes
this limitation.  The basic idea is to exploit the restoration of
chiral symmetry at short distances in the renormalized theory. This is
achieved by constructing finite improved operators by matching lattice
and continuum correlators at short distances. The matching is
implemented by requiring the vanishing of chirality violating form
factors in the correlation functions. This procedure is justified by
the following two observations.  Firstly, at short distances,
renormalized perturbation theory becomes chirally invariant (explicit
chiral symmetry breaking effects induced by the regularization are
reabsorbed by imposing the validity of the Ward Identities of chiral
symmetry, while violations from the non-vanishing of quark masses
disappear at short distances).  Secondly, contributions due to the
spontaneous breaking of chiral symmetry, which are absent in
perturbation theory, vanish in this region.  So both effects decrease
as we go to short distances.  
In all the cases of interest, including
the matrix elements of quark bilinear operators considered in this
letter, the terms which violate chirality vanish either quadratically
with the quark mass, i.e. as $m^2|x|^2$, where $|x|$ is the Euclidean
distance, or linearly as $m \lqcd^3 |x|^4$. On the other hand the
lattice artefacts vanish as $ma$. It is the different short-distance
behaviour of the $O(ma)$ terms which allows us to eliminate them. Note
that in this paper we are proposing a method for the determination of the
coefficients of the $O(ma)$ terms. For this reason the 
quark mass is a free parameter, which can be chosen to optimise the
procedure and does not necessarily coincide with any physical quark mass.
For example we may envisage using a mass of the order of a few hundred 
MeV to extract the coefficients, and use these coefficients in studies
of heavy quark physics.

The reason why mass corrections, other than the lattice artefacts,
vanish either as $m^2|x|^2$ or as $m \lqcd^3 |x|^4$, can be easily
understood. The chiral structure of QCD is such that contributions to
gauge invariant correlation functions which arise in perturbation
theory must contain an even number of mass insertions. For this reason
the short distance perturbative corrections vanish at least as
$m^2|x|^2$. On the other hand contributions due to spontaneous symmetry
breaking require the insertion of the mass operator, $m \bar q q$. 
By dimensional counting this implies that the corrections are of the 
form  $m \lqcd^3 |x|^4$.

Before presenting the technical details of the method, we explain what
we mean by ``short distances" throughout this paper. In the continuum
we simply mean that $|x|\lqcd\ll 1$, where $|x|$ is the Euclidean
distance. On the lattice, in addition, we require that $|x|/a$ is
sufficiently large so as to avoid lattice artefacts ($a$ is the
lattice spacing). The existence of a window, such that $a \ll |x|\ll
\lqcd^{-1}$, requires that we have a sufficiently small lattice
spacing. 

To implement a full $O(a)$ improvement away from the chiral limit,
further conditions should be satisfied. Indeed, it is necessary to
vary the lattice spacing $a$, distance $|x|$ and quark mass $m$ in
such a way that the chirality violating terms of $O(m^2|x|^2)$ and
$O(m \lqcd^3 |x|^4)$ are small enough to isolate the $O(ma)$ artefacts
in the correlation functions. We postpone the detailed description of
these conditions until subsection~\ref{subsec:scalar}, where a
specific example is discussed. For the remainder of the paper we will
assume that $|x|$ and $m$ are chosen so that the terms of
$O(m^2|x|^2)$ are much larger than those of $O(m \lqcd^3 |x|^4)$
(this condition is straightforward to satisfy provided the lattice
spacing is sufficiently small).

We wish to stress that the requirement of the existence of the window
$a \ll |x|\ll \lqcd^{-1}$ is common to all renormalization
schemes implemented using computations at a single value of $\beta$
(=$6/g_0^2$). In principle the constraint $|x|\ll\lqcd^{-1}$
can be avoided by using a sequence of lattices of decreasing $a$, and
correspondingly of decreasing physical volumes. The idea is to
determine the normalization constants by imposing renormalization
conditions on a lattice with a very small spacing. On such a
fine-grained lattice, very small distances $|x|$ are possible (so the
renormalization scale, in momentum space, can be chosen to be very
large) but the small physical size precludes the calculation of
hadronic correlation functions. It is therefore necessary to determine
the equivalent normalization constants for the larger, but coarser,
lattice, on which physical matrix elements are computed. This is
achieved by increasing the lattice spacing $a$ and matching the
normalization constants at fixed $|x|$, and then by increasing $|x|$
and matching at fixed $a$. This procedure is repeated until the
required value of $a$ is reached. This method has been proposed within
the context of the Schr\"odinger functional approach in
ref.~\cite{Luscher2}, but can also be applied here, and is illustrated
with an example in the appendix.

In ref.~\cite{MPSTV} a method of non-perturbative renormalization was
proposed, based on the imposition of renormalization conditions on
Green functions of operators between quark and gluon states in a fixed
gauge. It is possible to generalize this procedure to implement
improvement, exploiting the chiral behaviour of Green functions at
large momenta, but this adds a number of theoretical complications. In
a fixed gauge the basis of higher dimensional operators which must be
added to the bare lattice ones is larger than with the gauge invariant
method proposed here, particularly when computing off-shell Green
functions. This basis is restricted by BRST-invariance rather than
gauge invariance, and includes operators which vanish by the equations
of motion, but which are not themselves BRST-invariant~\cite{dawson}. A
detailed description of the procedure applied to quark bilinear
operators will be presented in ref.~\cite{dawson}, together with an
exploratory computation of all the subtraction coefficients.

The use of chiral Ward identities to impose constraints on the
normalization constants of operators is a well established
technique~\cite{wis}. In particular, for Wilson and tree-level improved
theories, they provide the normalizations of the vector and
axial-vector currents, the ratio of the normalizations of the scalar
and pseudoscalar densities and the values of the finite coefficients
describing the mixing of lattice operators of different chiralities
(this mixing is present due to the breaking of chirality by the Wilson
term).  Chiral Ward identities also provide important constraints for
the determination of the operators necessary for full $O(a)$
improvement~\cite{Luscher2,Luscher3} and, where possible, we propose 
to exploit them to determine some of the coefficients. As explained in
some detail in section~\ref{sec:wis}, however, it is not possible 
to determine all the improvement coefficients by the use of the Ward
identities only.

One can also attempt to construct an improved action and operators by
plotting the behaviour of the matrix elements of these operators as a
function of the lattice spacing, and fixing the improvement
coefficients by requiring that there are no linear terms in $a$
(including possible logarithmic factors). Such a procedure
is impracticable, however, at least without a considerable amount of
additional input based on the techniques described in
refs.~\cite{Luscher2, Luscher3} and in this letter.

The plan of the remainder of this letter is as follows. In the next
section we briefly review the construction of the improved action and
the determination of the mass dependence of the bare coupling constant
in the unquenched theory. In section~\ref{sec:wis} we explain why the
chiral Ward identities cannot be used to determine all the
coefficients needed to construct improved lattice composite
operators. The description of our method, and its application in the
construction of improved lattice quark bilinear operators for quarks
with non-zero masses is presented in
section~\ref{sec:impcomp}. Finally section~\ref{sec:concs} contains
our conclusions and a discussion of future prospects.

\section{Improvement of the action}
\label{sec:csw}

The Wilson fermion action ($S_{{\rm Wilson}}$) is improved by the
addition of the SW operator~\cite{sw},
\begin{equation}
S = S_{\rm gauge} + S_{\rm Wilson} + a \int d^4x\ \OSW 
\label{eq:swclover}\end{equation}
where $S_{\rm gauge}$ is the gauge action and 
\begin{equation}
\OSW = - \frac{i}4 c_{SW}\ \sum_{\mu\nu} \qbar \sigma_{\mu\nu}
F_{\mu\nu} q \, ,
\end{equation}
with $\sigma_{\mu\nu} = (i/2)[\gamma_\mu,\gamma_\nu]$.  Here and in the
following we use continuum notation to refer to lattice quantities. A
method for the non-perturbative determination of $c_{SW}$ was presented
and exploited in refs.~\cite{Luscher2,Luscher3}.

In order that physical quantities should approach their continuum
values with deviations of only $O(a^2)$, we need, in addition to
choosing the correct value of $c_{SW}$, to adjust the bare gauge
coupling $g_0$ and the bare mass $m_0$ in a way which depends on the
renormalized quark mass $m$~\cite{Luscher2}. To see this, consider the
renormalized coupling constant $g(\mu^2)$ defined, for example, as the
three gluon vertex (in some fixed gauge) at large virtualities
characterized by the scale $\mu^2$. The corresponding renormalization
constant $Z_g$, defined by
\begin{equation}
g(\mu^2) = Z_g\, g_0(a)\ ,
\label{eq:zgdef}\end{equation}
contains terms of $O(ma)$ from diagrams which include quark loops,
which must therefore be removed for improvement of unquenched lattice
QCD. Following ref.~\cite{Luscher2} we absorb this linear term into a
redefinition of the bare coupling constant $\tilde g_0$,
\begin{equation}
\tilde g_{\,0}^{\,2} = g_0^2\,(1+b_gma)\ ,
\label{eq:g0tildedef}\end{equation}
so that $Z_g$ has no remaining linear term in $ma$.

Thus, for each value of the bare quark mass, we need to determine 
$\tilde g_0$~\footnote{ See ref.~\protect\cite{Luscher2} for a method
based on the use of Schr\" odinger Functionals.}. For simplicity we
take the quarks to be degenerate, so that there is only one mass
parameter $m_0$; the generalization of the following discussion to the
non-degenerate case is straightforward. We can imagine, at least in
principle, proceeding as follows:
\begin{itemize}
\vspace{-10pt}
\item Choose a value of $g_0$, compute a short-distance  physical
quantity $Q$ at several values of the quark mass at
this value of $g_0$ and extrapolate the results to the chiral limit. An
example of a suitable quantity is the derivative of the potential
between a static quark and antiquark ($V^\prime(r)$) at a separation of
a fixed number of lattice spacings ($r=na$), satisfying $r\lqcd\ll 1$
and $rm\ll 1$. We denote the result by $Q(n, g_0, 0)$, where the final
argument represents the fact that this value corresponds to $m=0$.
\item The quantity $Q$ must be such that its linear dependence on the 
mass is entirely due to lattice artefacts, and that these $O(ma)$
effects can be removed by fixing $c_{SW}$ and $b_g$ only. For physical
quantities, such as decay constants and form factors, which are related
to matrix elements of composite operators, there are further mass
dependent terms of $O(ma)$ in their renormalization constants. It is
therefore not possible in general to use these quantities to determine
$b_g$ (with the exception of the matrix elements of the vector current, 
see subsection~\ref{subsec:vector} below).
\item We now choose a value of $m_0$ and recompute $Q$ at different
values of the bare coupling until we obtain the same result, i.e.
until we find a coupling $\tilde g_0$ such that $Q(n, \tilde g_0, m)
=Q(n, g_0, 0)$. 
In this way we obtain $\tilde g_0$ for each $m_0$. As
discussed in the introduction the
result is independent of $n$, up to quadratic corrections of order
$(mr)^2$. 
\item Finally, we need to determine the values of the lattice spacing
corresponding to $\tilde g_0(a)$ and of the bare quark mass ($m_0$) of
the physical quark.  This can be done by simulating with a series of
values of $\{m_0, \tilde g_0(m_0)\}$ and requiring that the computed
results for two physical quantities, for example the masses of two
hadrons, agree with their experimental values.
\end{itemize}

The technique described above needs considerable computing resources,
since it requires new simulations at a number of different values of
$\beta$ in order to map out the behaviour of $\tilde g_0$ with $m_0$.
At this stage we are not very concerned about this, since it will be
some time before an accurate determination of $b_g$ is actually
required. It will only be needed when precise unquenched computations
become possible ($b_g$ can be taken to be zero in quenched
calculations).  Even then, however, the effects of improvement (beyond
e.g. tree-level) will be most significant for heavy quarks, for which
the effects of quark loops are expected to be strongly suppressed.  We
also note that in first order perturbation theory $b_g$ is small, $b_g
\simeq 0.012\,N_f\,g_0^2$, where $N_f$ is the number of quark
flavours~\cite{ss}~\footnote{ The one-loop results for other
improvement coefficients can be found in ref.~\protect\cite{sint}.}.
Thus, for the purposes of this letter, we simply note that a
non-perturbative computation of $b_g$ is possible in principle and
postpone any detailed consideration of the most practical way of
determining it. We focus instead on the more immediate problem of
determining improved composite operators, thus, in particular,
removing the $O(ma)$ effects due to heavy valence-quarks.

For the remainder of this letter we assume that $c_{SW}$ is known, and
that, as the bare quark mass $m_0$ is varied, the bare coupling constant
has been readjusted to the corresponding value $\tilde g_0(m_0)$.

\section{Chiral Ward identities}
\label{sec:wis}

Before describing our method for the construction of improved lattice
composite operators, we explain why it is not possible to use chiral
Ward identities to obtain all the necessary coefficients. As an
example, consider a lattice quark bilinear operator $\bar q \Gamma q$,
where $\Gamma$ is one of the Dirac matrices, and we have suppressed the
flavour indices. The improved lattice operator takes the form
\begin{equation}
\widehat O_\Gamma = Z_\Gamma\left(\bar q \Gamma q + a c_\Gamma O_{4,\Gamma}
\right)\ ,
\label{eq:onshell}\end{equation}
where the $O_{4,\Gamma}$ are possible dimension 4 operators with the
same quantum numbers as $O_\Gamma$ (which will be exhibited explicitly
in the following section; there are no such operators for the scalar
or pseudoscalar densities). The operators on the right-hand side of
eq.(\ref{eq:onshell}) are the bare lattice operators.  $\widehat O$ is
constructed so that its matrix elements between physical states have
discretization errors only of $O(a^2)$. Following
ref.~\cite{Luscher2}, we parametrize the mass dependence in the
overall renormalization constants by
\begin{equation}
Z_\Gamma(m) = Z_\Gamma(0) (1 + b_\Gamma\, ma)\ . 
\label{eq:zgamma}\end{equation}
The point which we wish
to stress here is that the Ward identities do not determine
the $b_\Gamma$'s. In the remainder of this section we briefly explain
the reasons for this limitation and in the following one we
propose a new method for the determination of the $b_\Gamma$'s.

Consider the following continuum Ward identity
\begin{equation}
\left<\left\{\int_R d^4x \left[
2 m P(x) - \del_\mu A_\mu(x)\right]\,O(y)-\delta O(y)\right\}\,
O_1(z_1)\,O_2(z_2)\cdots\right>\,=\,0\ ,
\label{eq:wi1}\end{equation} 
where $P$ and $A_\mu$ are a pseudoscalar density and axial current
respectively, and $O$, $O_1$, $O_2$ etc. are composite operators,
which, for the purposes of this letter, we can assume to be gauge
invariant. We are free to make any convenient choice of the operators
$\{O_1, O_2,\cdots\}$.  For simplicity of presentation we suppress the
flavour indices on all the operators. The integral is over a region
$R$ which includes the point $y$, but excludes the points
$z_i$. $\delta O$ is the axial variation of $O$.  The improved
lattice version of the operator $2 m P(x) - \del_\mu A_\mu(x)$ is
obtained (up to an overall normalization) by requiring it to satisfy
Ward identities of the form (up to terms of $O(a^2)$)
\begin{equation}
\langle\,\left[\,
2 m P(x) - \del_\mu A_\mu(x)\,\right]\,
O_1(z_1)\,O_2(z_2)\cdots\,\rangle\,=\,0\ ,
\label{eq:wi0}\end{equation} 
where $x$ is separated from the other points $z_i$. It might therefore
be hoped that by subsequently imposing the identity in
eq.(\ref{eq:wi1})  it would be possible  to determine the improved
forms of $O$ and $\delta O$ (up to an overall normalization factor).
This is not so, however, since in order to satisfy this identity up to
terms of $O(a^2)$, using lattice operators, we must add irrelevant
operators to $P(x)$ and $O(y)$ which ``vanish'' by the equations of motion,
and which therefore do not contribute to physical matrix elements. For
example, the improved form of $O_\Gamma$ in off-shell matrix elements
takes the form
\begin{equation}
\widehat O_\Gamma = Z_\Gamma\left(\bar q \Gamma q 
+ a c_\Gamma^\prime\,\bar q\left[\Gamma (\rDslash + m_0)
+(-\lDslash + m_0)\Gamma\right]q + a c_\Gamma O_{4,\Gamma}\right)\ ,
\label{eq:offshell}\end{equation}
where $\rDslash + m_0$ is the lattice version of the fermion matrix
appearing in eq.(\ref{eq:swclover}). The terms proportional to the
new coefficients $c_\Gamma^\prime$ contribute to the identity in
eq.(\ref{eq:wi1}) in the region where $P(x)$ comes into contact with
$O(y)$~\footnote{ Note that the integral of $\del_\mu A_\mu$ in
eq.(\ref{eq:wi1}) gives a surface integral, and so produces no contact
terms.}. Thus, because of these ``contact terms", there are extra
coefficients, the $c_\Gamma^\prime$'s, to determine, and in fact it is
not possible to obtain the $c_\Gamma^\prime$'s and the $b_\Gamma$'s
separately using the Ward identities (and so it is not possible to
obtain the $b_\Gamma$'s which are needed for the evaluation of
physical matrix elements).  

\par We now justify this claim in more detail. Using the expressions
(\ref{eq:offshell}) for the operators $P, A_\mu, O$
and $\delta O$ in eq.(\ref{eq:wi1}), we find that the term containing
the rotated lattice operator $\delta O$ is
\begin{equation}
-Z_AZ_O\left[\frac{Z_{\delta O}}{Z_AZ_O} - 2 \rho a (c_P^\prime + c_O^\prime)
\right]\,\langle \delta O(y)\,O_1(z_1)\,O_2(z_2)\cdots\rangle\ ,
\label{eq:cantdo}\end{equation}
where $\rho$ is computed directly from the lattice correlation functions
in eq.(\ref{eq:wi0}):
\begin{equation}
2\rho = \frac 
{\left<\,\del_\mu A_\mu(x)\,
O_1(z_1)\,O_2(z_2)\cdots\right>}{
\left<P(x) \,O_1(z_1)\,O_2(z_2)\cdots\right>}\ .
\label{eq:rhodef}\end{equation}
In eq.(\ref{eq:rhodef}), $x$ is separated from the $\{z_i\}$, and 
$\rho$ is independent of the choice of the operators $\{O_i\}$ and of
all the coordinates. Since the term proportional to $\rho$ in
eq.(\ref{eq:cantdo}) is multiplied by a factor of $a$, it is
sufficient to take any convenient, not necessarily improved, lattice
pseudoscalar density $P$ or axial current $A_\mu$ in
eq.(\ref{eq:rhodef})~\footnote{ In this letter we have not needed to
define an improved quark mass. However, comparing equations
(\protect\ref{eq:wi0}) and (\protect\ref{eq:rhodef}), leads us to
write $\rho = m Z_P/Z_A$. By computing $\rho$ using improved operators
$A_\mu$ and $P$, whose construction is explained in
section~\ref{sec:impcomp}, the corresponding mass $m$ is indeed a
renormalized improved quantity.}. 
From eq.~(\ref{eq:cantdo}) we see that the
$b_\Gamma$'s and the $c_\Gamma^\prime$'s appear in the combination:
\begin{equation}
\frac{Z_{\delta O}(0)}{Z_A(0)Z_O(0)}\,\left\{\,1 + (b_{\delta O} -b_O
-b_A)\, ma\,\right\}
-2 \rho a (c_P^\prime + c_O^\prime)\ ,
\label{eq:cantdo2}\end{equation}
where the $b_\Gamma$'s are defined in eq.(\ref{eq:zgamma}).
For each bilinear $O$, the $O(a)$
part of the Ward Identity in eq.(\ref{eq:wi1}) allows one
to obtain the combination of $O(a)$ that appear in
eq.(\ref{eq:cantdo2}).  Thus one only obtains a linear combination of
the $b_\Gamma$'s and the $c^\prime_\Gamma$'s, and even by considering
all possible bilinears it is not possible to determine any of the
$b_\Gamma$'s separately.

Despite this shortcoming, chiral Ward identities should be part of any
improvement program; they provide a determination of the $c_\Gamma$'s as
well as many checks of the numerical results for the remaining
improvement coefficients.

\section{Improvement of bilinear quark operators}
\label{sec:impcomp}

In this section we propose a new method for the determination of the
improvement coefficients in general, and the $b_\Gamma$'s in
particular.  We illustrate the method, by discussing the
construction of improved quark bilinear operators. We
consider only flavour non-singlet bilinears, but, unless specifically
required for the discussion, we drop the flavour indices.

\subsection{Pseudoscalar and scalar densities}
\label{subsec:scalar}

The renormalized, improved pseudoscalar density has the general form
\begin{equation}
\widehat{P}(x) \equiv Z_P P(x) \, , 
\label{eq:phat}\end{equation} 
where the renormalization constant $Z_P\equiv Z_P(m)=Z_P(0)(1+b_P \, ma
+ \cdots)$ is, in general, a function of the quark mass and $P(x)=
\qbar(x) \gamma_5 q(x)$ is the bare lattice pseudoscalar density. We
start by constructing the two-point correlation function $\widehat
G_{PP}(x)=\langle \widehat{P}(x) \widehat{P}(0) \rangle$ and by
studying it in the limit $\vert x \vert \to 0$. By ``$\vert x \vert \to
0$" we mean ``short distances" in the sense explained in the
Introduction. In this region, we expect on the one hand to be able to
use the operator product expansion, keeping only the leading, most
singular term, and yet on the other hand to neglect contributions from
contact terms.

At any given value of $m$, provided $m a \ll 1$, we determine
$Z_P$ by imposing, at short distances, the following renormalization
condition
\begin{eqnarray} \widehat G_{PP}(x) &=&
 Z_P^2(m) \langle {P}(x)  {P}(0) \rangle \nonumber \\&=&
\langle \hat{P}(x)\hat{P}(0)\rangle_{{\rm cont}} \mbox{\ \ as} \,\,\,
\, |x| \to 0 \, .
\label{eq:zp} \end{eqnarray}
$G_{PP}(x)\equiv\langle {P}(x) {P}(0) \rangle$ is the bare lattice
pseudoscalar-pseudoscalar correlation function, computed by numerical
simulation at the quark mass $m$.  In eq.(\ref{eq:zp})
$\langle\hat{P}(x) \hat{P}(0) \rangle_{{\rm cont}}$ is the
pseudoscalar-pseudoscalar correlation function computed in perturbation
theory in some continuum renormalization scheme, e.g. in the
$\overline{MS}$ or the $RI$ schemes~\cite{MPSTV} at a renormalization
scale $\mu$. A convenient choice is $\mu\sim 1/|x|$.
At short distances $\langle\hat{P}(x) \hat{P}(0) \rangle_{{\rm cont}}$ 
is independent of the quark mass, up to subasymptotic corrections
vanishing as $m^2|x|^2$.  This shows that we can determine the $O(am)$
correction to $Z_P(m)$ using only gauge invariant quantities.

\par  An alternative method to determine $Z_P$ is the following. 
We define, at short distances,
\begin{equation} 
R_P= (1 +  b_P m a)^2  = \frac{G_{PP}(x)|_{m=0}}{G_{PP}(x)|_{m}}\, ,
\label{eq:rpdef}\end{equation}
where $G_{PP}(x)|_{m=0}$ is the bare lattice correlator $G_{PP}$
extrapolated to the chiral limit. Thus a computation of the ratio
$R_P$ provides us with a determination of $b_P$. In addition, at short
distances $R_P$ is independent of $x$, which provides us with a set of
consistency conditions. 

The leading physical chirality-violating
correction to $R_P(x)$ is of the form (up to logarithmic corrections)
$m^2 |x|^2$. In order to be able to extract $b_P$ we require $m^2
|x|^2\ll ma$, which at fixed $a$ imposes constraints on the allowed
values of $m$ and $|x|$. In practice, for values of the lattice
spacing currently used in simulations ($a^{-1} = O(2$--4 GeV$)$),
these conditions can be implemented without difficulty~\footnote{ At
short distances, the terms of $O(m \lqcd^3 |x|^4)$ are naturally
negligible compared to those of $O(m^2 |x|^2)$ provided that the mass
$m$ is not much smaller than $\lqcd$.}. In addition, one can estimate
the $O(m^2 |x|^2)$ terms by studying the short-distance dependence of
$R_P(x)$ on $x$ at fixed $m$, and try in this way to improve the
precision in the determination of $b_P$. These considerations also
apply to all the bilinears considered below. 

The above is a good example illustrating the necessity of knowing
$b_g$ (or equivalently, of knowing $\tilde g_0(a)$ for each value of
the quark mass).  The numerator and denominator on the right-hand side
of eq.(\ref{eq:rpdef}) have to be evaluated at the same values of the
coordinates $x$ (in physical units).  As explained in
section~\ref{sec:csw}, in order to satisfy this requirement (up to
$O(a^2)$ corrections), the bare coupling constant has to be varied as
a function of the mass as in eq.(\ref{eq:g0tildedef}).

\par Having determined the mass dependence of $Z_P(m)$, $Z_P(0)$ can be
obtained for example, by imposing renormalization conditions on the
pseudoscalar density in the chiral limit using the Schr\" odinger
functional techniques~\cite{Luscher3} or by determining $Z_P(m)$ on
quark states, as suggested in ref.~\cite{MPSTV}, and then by
extrapolating the results for $Z_P(m)$ to the chiral limit. In the
latter case we can ignore completely the extra terms coming from
operators which vanish by the equation of motions, or from non-gauge
invariant operators, since the systematic error due to these terms
disappears as $m \to 0$~\cite{dawson}.

\par The determination of the renormalization constant of the scalar
density, $Z_S$, proceeds in exactly the same way. The ratio of the
renormalization constants $Z_S(0)/Z_P(0)$, which is a finite quantity, can be
determined in the standard way using chiral Ward Identities~\cite{wis}.

\subsection{The Vector Currents}  
\label{subsec:vector}

The improved vector current has the general form
\begin{equation}
\widehat{V}_\mu(x) = Z_V  V_\mu(x) \ ,
\label{eq:vhatdef}\end{equation}
where
\begin{equation}
V_\mu(x) =
\qbar(x)\gamma_\mu  q(x)
+ a c_V \sum_\nu i \del_\nu \left[\qbar(x)\sigma_{\mu\nu}  q(x)\right]
 \, ,
\label{eq:vmu} \end{equation}
and the operators on the right-hand side of (\ref{eq:vmu}) are the 
bare lattice ones.
In this case, improvement requires the determination of two constants
($Z_V$ and $c_V$), one more than for the pseudoscalar and scalar
densities. The necessity of the term proportional to $c_V$ for improvement
has been pointed out and discussed in refs.~\cite{Luscher2,msv}. 
For any choice of the quark mass, $Z_V\equiv Z_V(m)=Z_V(0)(1
+ b_V a m)$ can be immediately fixed by requiring that the forward
matrix element of the current is correctly normalized,
i.e. that 
\begin{equation}
\langle p  \vert \widehat{V}_\mu \vert p \rangle= Z_V(m)
\langle p  \vert {V}_\mu \vert p \rangle= 2 p_\mu \, .\end{equation}
This does not require the knowledge of $c_V$ since the forward matrix
element of $\del_\nu \left[\qbar(x)\sigma_{\mu\nu} q(x)\right]$ is
zero. We then determine $c_V$ by studying the short-distance behaviour
of the correlation function
\begin{equation}
\widehat G_{VV}(x)= 
\langle 0 \vert \widehat{V}_\mu(x) \widehat{V}_\nu(0) \vert 0 \rangle 
\, .\end{equation} 
$c_V$  is to be chosen to remove the residual dependence of $\widehat
G_{VV}(x)$ on $ma$ as $\vert x \vert \to 0$. 

It is also possible to exploit the chiral Ward identities (in the
massless limit) to determine $c_V$~\footnote{ This is the method
currently being used by the authors of
ref.~\protect\cite{guagnelli}.}.  For example, having determined $Z_V$
from the normalization of the charge operator, and anticipating that
we can obtain the improved axial current as described in
subsection~\ref{subsec:axial} below, the Ward identity (see
eq.(\ref{eq:wi1})\,)
\begin{equation}
f^{abc} \sum_{\vec x}\langle\, \left(\, \hat{A}_0^a(t_2,\vec
x) - \hat{A}_0^a(t_1, \vec x) \,\right)\, \hat{A}_i^b(y)\,
V_{i,{\mathrm source}}^c(0)\,\rangle = i\,N_f\,\langle\,
\hat{V}_i^c(y)\,V_{i,{\mathrm source}}^c(0) \,\rangle
\label{eq:wiforcv}\end{equation}
provides an equation for $c_V$. In eq.(\ref{eq:wiforcv})
$V_{i,{\mathrm source}}^c(0)$ is any vector current (not necessarily
improved), we have assumed that $0< t_1< y_0< t_2$, and a sum over the
spatial index $i$ is implied. It has been convenient to exhibit the
flavour indices $\{a,b,c,\cdots\}$ explicitly~\footnote{ We use a
notation in which, for example, $P^a(x) = \bar\psi(x)\gamma^5\tau^a
\psi(x)$, with the generators of the flavour symmetry normalised to
${\mathrm Tr}(\tau^a\,\tau^b)=\frac{1}{2}\delta^{ab}$.  We present the
results for an $SU(N_f)$ flavour symmetry, and $f^{abc}$ are the
corresponding structure constants, $[\tau^a,\tau^b]
=if^{abc}\tau^c$.}.  By varying the parameters in
eq.(\ref{eq:wiforcv}) ($t_{1,2}, y$ and the form of $V_{i,{\mathrm
source}}^c(0)$\,) we obtain a series of consistency conditions.

We stress that when using Ward identities, such as that in 
eq.(\ref{eq:wiforcv}), we do not require ``short distances". Indeed for
$|y|\to 0$ in eq.(\ref{eq:wiforcv}), the contribution from the term
containing $c_V$ vanishes (i.e. is of $O(a^2)$), and hence cannot be
determined. To see this, note that this term is proportional to
\begin{equation}
ac_V\langle\,\bar q(y) \sigma_{\mu\nu}\tau^c q(y) \,V_{i,{\mathrm
source}}^c(0)\,\rangle\ , 
\label{eq:cvwi}\end{equation}
which vanishes at short distances by chirality (on the lattice the
correlation function in eq.(\ref{eq:cvwi}) is of  $O(a)$ because of the
Wilson term, leading to a contribution of $O(a^2)$). Thus, in order to
use the Ward identity to determine $c_V$ we need to work at distances
$y\gtrsim 1/\lqcd$, where the non-perturbative chiral symmetry breaking terms
make the correlation function in eq.(\ref{eq:cvwi}) non-vanishing.

If the Ward identities are used to determine $c_V$, then rather than
using the independence of $\widehat{G}_{VV}(x)$ of the quark mass to
determine $c_V$, we can use it instead to determine $\tilde g_0(m)$
($b_g$). Thus, we see that there are many possibilities to
(over-)determine the set of parameters $\{ b_g, Z_V(0), b_V, c_V\}$.

\subsection{The Axial Currents}
\label{subsec:axial}

The improved form of the axial current is
\begin{equation}
\hat{A}_\mu(x) = Z_A\,  A_\mu(x) 
\label{eq:ahatdef}\end{equation}
where
\begin{equation}
A_\mu(x) = 
\qbar(x)\gamma_\mu \gamma_5 q(x)
+ a c_A \del_\mu \left[\qbar(x) \gamma_5 q(x)\right] \, ,
\label{eq:amu}\end{equation}
and the operators on the right-hand side of (\ref{eq:amu}) are bare
lattice ones.  As in the case of the vector current, there are two
constants ($Z_A$ and $c_A$) to determine, but now there is no
normalization condition which would allow us to determine $Z_A$
without knowledge of $c_A$.  We start therefore with the determination
of $c_A$.  Following~\cite{Luscher2,Luscher3} we exploit the fact that
$\del_\mu \hat{A}_\mu(x)$ is proportional to $P(x)$, and compute
\begin{equation}
R_A(x,c_A)=\frac{\langle\,\del_\mu\hat{A}_\mu(x)\,P(0)\rangle}
{\langle P(x)\,P(0)\rangle}\ .
\label{eq:radef}\end{equation}
$c_A$ is determined, together with $c_{SW}$, by requiring that $R_A$
is independent of $x$ (up to terms of $O(a^2)$ which we are
neglecting).  $R_A$ is also independent of the form of the
pseudoscalar operator at the origin (e.g. $P(0)$ could be replaced by
some extended operator in both the numerator and denominator of
eq.(\ref{eq:radef}); see also the discussion around
eqs.(\ref{eq:rhodef}) and (\ref{eq:wiforcv})\,). This provides further
constraints which can be exploited in the simultaneous determination
of $c_A$ and $c_{SW}$~\cite{Luscher2}.

The determination of $Z_A$ can now be achieved in a number of
ways. For example, one can start by calculating $Z_A$ in the chiral
limit by imposing a chiral Ward identity~\cite{Luscher3}, using the
improved operators $\hat{V}$ and $\hat {A}$. We will see below that in
this case we do not need to know the values of $Z_V$ and $c_V$. For
example, consider the identity
\begin{equation}
f^{abc}\sum_{\vec x, \vec y}\langle\, 
P^e(z)\left(\, \hat{A}^a_0(t_2,\vec x) -  \hat{A}^a_0(t_1,
\vec x) \,\right)\, \hat{A}_0^b(y)\, P^d(0)\,\rangle
=i\,N_f\,\sum_{\vec y}\langle\,
P^e(z)\,\hat{V}_0^c(y)\,P^d(0)\,\rangle\ ,
\label{eq:wiforza}\end{equation}
where $0<t_1<y_0<t_2<z_0$ and again it has been convenient to exhibit
the flavour indices $\{a,b,c,\cdots\}$ explicitly. Any convenient form
for the lattice pseudoscalar densities, $P$, can be used in
eq.(\ref{eq:wiforza}); indeed the pseudoscalar densities could also
be replaced by other operators. Both sides of eq.(\ref{eq:wiforza})
can be simplified. Current conservation implies that the left-hand
side is independent of $t_{1,2}$ and $y_0$ (provided that the above
time-ordering is preserved) and hence that the two terms are equal
(note that as we are summing over $\vec x$ and $\vec y$, we can, for
example, change variables $\vec x\to\vec y$ in the second term).
Combining this with the normalization of the charge operator on the
right-hand side we obtain
\begin{equation}
f^{abc}f^{cde}
\sum_{\vec x, \vec y}
\langle\, P^e(z)\,\hat{A}^a_0(x)\, \hat{A}_0^b(y)\, P^d(0)\,\rangle
= - \frac{N_f^2}{2}\,\langle P^c(z)\, P^c(0)\,\rangle\ ,
\label{eq:wiforza2}\end{equation}
where $0<y_0<x_0<z_0$.
For each set of values of $\{x_0, y_0, z_0\}$ satisfying this ordering,
and for each choice of $\vec z$,
eq.(\ref{eq:wiforza2}) is an equation for the one unknown
$Z_A(0)$ ($c_A$ having been determined as
described above). 
Finally, the values of $Z_A(m)=Z_A(0)( 1 + b_A ma )$
away from the chiral limit can be obtained from the condition that at
short distances the correlation function $\langle\, \hat{A}_\mu(x)
\hat{A}_\nu(0)\rangle$ is independent of the quark mass,
\begin{equation}
\langle\, \hat{A}_\mu(x)\hat{A}_\nu(0)\rangle|_{m} = \langle\,
\hat{A}_\mu(x) \hat{A}_\nu(0)\rangle|_{m=0}\ .
\label{eq:zamneq0}\end{equation}
For any value of the quark mass, eq.(\ref{eq:zamneq0}) represents an
equation for $b_A$. With this method, no perturbative calculations,
either on the lattice or in the continuum are necessary.

Alternatively one can calculate the correlation function $\langle\,
A_\mu(x) A_\nu(0)\,\rangle$ at short distances in continuum
perturbation theory (here $A_\mu$ is the continuum axial current and,
at short distances, the correlation function is independent of the
mass) and determine $Z_A(m)$ by imposing that $\langle\, \hat{A}_\mu(x)
\hat{A}_\nu(0)\,\rangle$ is equal to this perturbative result. Finally,
as done for the pseudoscalar density, we can compute  $Z_A$ on quark
states, for different masses, and obtain $Z_A(0)$ by extrapolating the
result to the chiral limit. This gives the correct result provided the
renormalization scale is large enough and the renormalization
conditions are compatible with the Ward identities for the axial
current~\cite{MPSTV}.

\subsection{The Tensor bilinears}
Finally, we briefly sketch the corresponding analysis for the tensor
operator. The improved tensor operator has the form
\begin{equation}
\hat{T}_{\mu\nu}(x) = Z_T T_{\mu\nu}(x) 
\label{eq:thatdef}\end{equation}
where
\begin{equation}
T_{\mu\nu}(x) = i\,\qbar(x)\sigma_{\mu\nu} q(x)
+ a c_T (\del_\mu V_\nu - \del_\nu V_\mu) \,,
\label{eq:tmunu}\end{equation}
and the operators on the right-hand side of (\ref{eq:tmunu}) are bare
lattice ones.  Since we are neglecting terms of $O(a^2)$, we do not
need to include the $O(a)$ terms proportional to $c_V$ in the
expression for $V_\mu$ given in eq.(\ref{eq:vmu}).

Again we have two constants ($Z_T$ and $c_T$) to determine and, as in
the case of the axial current, we start with the determination of
$c_T$. This can be achieved by noting that the correlation function
$\langle\, \hat{T}_{\mu\nu}(x)\,\hat{V}_\rho(0)\, \rangle$ is zero at
short distances. This is because the tensor and vector currents have
different chiralities, and chirality is a good symmetry at short
distances. The vanishing of this correlation function represents an
equation for $c_T$. 

The coefficient $c_{T}$ can also be determined using Ward Identities
in the chiral limit. For example, consider the identity
\begin{equation}
\sum_{\vec x} \sum_{\mu\nu}
\epsilon_{\mu\nu\rho\sigma}
\langle\, \left(\, \hat{A}_0^a(t_2,\vec x)-\hat{A}_0^a(t_1,\vec x) \,\right)\, 
\hat{T}_{\mu\nu}^b(y)\, T_{\rho\sigma,{\mathrm source}}^c(0)\,\rangle
= 2 d^{abc}
\langle\, \hat{T}_{\rho\sigma}^c(y)\,T_{\rho\sigma,{\mathrm source}}^c(0)
\,\rangle \,,
\label{eq:wiforct}
\end{equation}
in which no indices are summed unless explicitly indicated,  and we
have assumed $0<t_1 < y_0 < t_2$.  This identity is valid for any form
of the tensor density at the origin; the simplest case is to assume
that $T_{\mu\nu,{\mathrm source}}=i \bar q \sigma_{\mu\nu} q$.  The set
of equations (\ref{eq:wiforct}) for different choices of $\rho$,
$\sigma$, $y$ and $T_{\mu\nu,{\mathrm source}}$, allow us to determine
$c_T$ as well as providing  many consistency checks.  Note that $Z_T$
appears as an overall factor on both sides of the equation,  and so
cannot be obtained from eq.(\ref{eq:wiforct}).

The overall normalization constant $Z_T$ can then be determined, for
example, by computing the correlation function $\langle
\hat{T}_{\mu\nu}(x)\,\hat{T}_{\rho\sigma}(0)\,\rangle$ at short
distances in perturbation theory (in some renormalization scheme) and
on the lattice and imposing that they are equal. $Z_T(m)$ is now the only
unknown in this equation and can therefore be obtained.

$Z_T$ can also be obtained in the chiral limit by imposing a
normalization condition on the matrix element of the tensor operator
between quark states. The mass dependence of $Z_T(m)$ is then
determined in the standard way, by requiring that the correlation
function $\langle \hat{T}_{\mu\nu}(x)\,\hat{T}_{\rho\sigma}(0)\,
\rangle$ is independent of the quark mass at short distances.

\section{Conclusions}
\label{sec:concs}

In this letter we have proposed a method for the construction of
improved lattice composite operators, containing quarks with non-zero
masses. The remaining discretization errors in matrix elements of
these operators are of $O(a^2)$. The method builds on the improvement
program of the {\sc Alpha} collaboration and extends it away from the
chiral limit. This is of particular importance in the study of the
decays of hadrons containing a heavy quark.  A detailed numerical
study is now necessary to determine the precision with which the
coefficients can be obtained; the results will be presented in a
future paper.

Our proposed method is based on the observation that chiral symmmetry
is restored at short distances. By imposing this requirement on
correlation functions of composite operators, we can determine the
coefficients which are needed to construct the improved operators.
For purposes of illustration, in this letter we have studied the
application of our method to quark bilinear operators; the approach
can, however, also be generalized to deal with more complicated
composite operators, such as four-quark operators relevant for studies
of the effective weak Hamiltonian.

We are also carrying out a pilot study of the determination of the
normalization coefficients by imposing normalization conditions
between quark and gluon states in a fixed gauge. Again the key
ingredient in the method is the restoration of chiral symmetry at
large momenta or short-distances. The basis of higher dimensional
operators which must be considered (and hence the number of
coefficients) is larger in this case, and the method will be presented
in detail in ref.~\cite{dawson}, together with the results of the
preliminary numerical study.

\section*{Acknowledgements}
C.T.S., G.M., G.C.R., M.Ta. and M.Te. acknowledge partial support from
EU contract CHRX-CT92-0051. M. Talevi thanks the Universit\`a di Roma
``La Sapienza'' where part of the work was carried out and
acknowledges INFN for partial support and EPSRC for its support
through grant GR/K41663 and PPARC for its support through grant
GR/L22744.  C.T.S. acknowledges PPARC for its support through
grant GR/J21569. G.M., G.C.R., M.Ta. and M.Te. acknowledge partial
support by M.U.R.S.T. S.S. thanks the University of Rome, ``La
Sapienza", for its hospitality, and the INFN for partial support. S.S.
was also partially supported by the U.S. Department of Energy grant
DE-FG03-96ER40956.

\section*{Appendix}

In this appendix we sketch the procedure which allows, in principle,
the evaluation of $Z_P$ away from the chiral limit, in a way which
obviates the need for the condition $|x|\ll\lqcd^{-1}$.  Following
ref.~\cite{Luscher2}, we start with a fine lattice with a spacing
which is so small that one can perturbatively match the lattice
results onto a continuum renormalization scheme with negligible errors
from higher order corrections, and, after successive iterations, we
end with a course lattice which is sufficiently large to allow the
computation of hadronic correlation functions.  For simplicity we
present the discussion in the quenched case only, which obviates the
need to determine $b_g$. The generalization to the unquenched case
introduces further technical complications, but the basic strategy is
the same. The steps are the following~\footnote{ In this appendix, we
exhibit explicitly the relevant arguments of $Z_P$, $G_{PP}$, $R_P$
etc. where needed.  Some of these arguments were implicit in the
earlier sections.}:
\begin{enumerate}
\item[i)] On a lattice with a very small spacing, $a_1$, and bare
coupling $g_1$, we use our method to determine $Z_P(g_1, \mu, m_1)$
for a given value of the quark mass $m_1$, and for a renormalization
scale $\lqcd\ll\mu\ll a_1^{-1}$. This condition
is readily satisfied since we are working on a very fine lattice.
We do this using two-point functions at distances $|x_1|\sim 1/\mu$.
\item[ii)] We evaluate $R_P(x_2, g_1, m_1)$ as defined in
eq.~(\ref{eq:rpdef}) and a second physical quantity $Q(x_2, g_1, m_1)$
for which the $O(a)$ effects can be removed by adjusting $c_{SW}$ only
(see section~\ref{sec:csw}). We choose $|x_2|$ to be larger than
$|x_1|$, e.g. a simple choice is $|x_2| = 2 |x_1|$.
\item[iii)] The next step is to obtain the same physical quark mass on
both lattices.  With a coupling constant $g_2$, corresponding to a
lattice spacing $a_2$ larger than $a_1$, e.g. $a_2 = 2 a_1$, we adjust
the mass of the quark until $Q(x_2, g_1, m_1)=Q(x_2, g_2, m_2)$ ($x_2$
is the same in physical units on both lattices). The calibration of
the lattice spacing is performed, as usual, in the chiral limit,
using, for example, $Q(x_2, g_1, 0) = Q(x_2, g_2, 0)$.
\item[iv)] We find the new value of $b_P$ by imposing the condition
$R_P(x_2, g_2, m_2) = R_P(x_2, g_1, m_1)$. 
\item[v)] At this point we only require $Z_P$ in the chiral limit.
This is obtained by imposing the condition $Z_P(g_2, \mu, 0)
G_{PP}(x, g_2, 0) = Z_P(g_1, \mu, 0) G_{PP}(x, g_1, 0)$.
\end{enumerate}
By iterating the steps ii) -- v) we can reach a large lattice (in physical
units). 

For the other bilinears, one first determines the subtraction constants,
$c_V$, $c_A$ and $c_T$ in the chiral limit, and then use the above procedure
to obtain the overall renormalization constants.

\end{document}